\documentclass{article}
\usepackage{authblk}
\usepackage{graphicx}
\usepackage{amsmath}
\usepackage{float}
\usepackage{subcaption}
\usepackage{pgffor}
\usepackage{xspace}
\usepackage{amsfonts}
\usepackage[numbers]{natbib}
\title{SwdFold:A Reweighting and Unfolding method based on Optimal Transport Theory}
\newcommand{\OmniFold}{{\sc OmniFold}\xspace}
\newcommand{\SwdFold}{{\sc SwdFold}\xspace}
\newcommand{\SwdRw}{{\sc SwdRw}\xspace}
\date{}

\author[1]{Chu-Cheng Pan}
\author[1]{Xiang Dong}
\author[1]{Yu-Chang Sun}
\author[1]{Ao-Yan Cheng}
\author[1]{Ao-Bo Wang}
\author[1]{Yu-Xuan Hu}
\author[1]{Hao Cai\thanks{hcai@whu.edu.cn}}
\affil[1]{School of physics and technology, Wuhan University, Wuhan, Hubei, China}

\begin{document}

\maketitle

\begin{abstract}
  High-energy physics experiments rely heavily on precise measurements of energy and momentum, yet face significant challenges due to detector limitations, calibration errors, and the intrinsic nature of particle interactions. Traditional unfolding techniques have been employed to correct for these distortions, yet they often suffer from model dependency and stability issues. We present a novel method, \SwdFold, which utilizes the principles of optimal transport to provide a robust, model-independent framework to estimate the probability density ratio for data unfolding. It  not only unfold the toy experimental event by reweighted simulated data distributions closely with true distributions but also maintains the integrity of physical features across various observables. We can expect it can enable more reliable predictions and comprehensive analyses as a high precision reweighting and unfolding tool in high-energy physics.
\end{abstract}

In high-energy physics experiments, achieving precision and accuracy in measurements is critical for conducting meticulous analyses. However, detectors face inherent resolution limitations, calibration errors, and issues such as the generation of secondary particles, which can skew the measurement of energy and momentum. 
Unfolding is a mathematical process used to correct distortions in data caused by instrumental effects(detector level data), enabling the recovery of the true distribution of physical events(particle level data). This task is complex due to the diversity and intricacy of the physical event distributions, compounded by systematic errors and uncertainties from the detectors. Unfolding technique is particularly crucial in accurately determining physical observables like invariant mass distributions, where it aids in precisely measuring mass values and widths\cite{hoecker1996svd_unfold}. 
Moreover, unfolding can handle complex physical phenomena such as jet measurements, efficiently recovering original distributions from data convoluted by instrumental effects\cite{andreev2022h1_omnifold}. 
Additionally, by mitigating dependencies on specific experimental setups, unfolding enables consistent comparisons across different experiments and facilitates the application of advanced theoretical tools, ensuring that findings remain relevant and comparable long after the initial analyses are conducted\cite{andreassen2020omnifold, cowan2002survey_unfolding}. 

Unfolding in high energy physics is traditionally approached by discretizing the probability distribution into bins and transforming the unfolding problem into matrix inversion\cite{adye2011roounfold}. The iterative Bayes' theorem utilizes repeated applications of Bayes' theorem to invert the response matrix, with regularization achieved by limiting the number of iterations based on sample statistics and binning, thus avoiding overfitting despite initial conditions set by the training truth rather than a flat distribution\cite{d1995bayes_unfolding}. On the other hand, Singular Value Decomposition (SVD), inverts the response matrix using singular value decomposition techniques, where regularization involves suppressing contributions from small singular values that signify high-frequency fluctuations, thus enhancing stability and accuracy\cite{hoecker1996svd_unfold}. Lastly, Tikhonov regularization, uses polynomial regularization to stabilize the inversion of matrices when mapping multi-dimensional distributions, with the optimal regularization parameter determined via an L-curve analysis\cite{schmitt2012tunfold}. 

To overcome the limitations associated with traditional methods, such as the need for binning and the restriction to single observables, a new approach, known as \OmniFold,  involve the iterative reweighting of a simulated event set with machine learning techniques has been introduced\cite{andreassen2020omnifold}. The reweighting process employs deep neural network-based classifiers to continuously estimate probability ratio and adjust weights within a simulation\cite{andreassen2020omnifold_reweight_theory}. The methodology can be conceptualized as an EM-type algorithm, which iteratively maximizes the likelihood estimate\cite{andreassen2021omnifold_cs}. However, while the EM algorithm is widely used for its convergence properties, it only guarantees convergence to a stationary point of the likelihood function\cite{kolouri2018swdgmm}. Studies indicate that the likelihood function may have multiple local maxima, some of which can result in significantly worse log-likelihood values compared to the global optimum\cite{amendola2016mlgmm, hoffmann2005unsupervised_gmm, jin2016local_maxima_ml}. Moreover, with random initialization, the EM algorithm tends to converge to these suboptimal critical points with high probability, making the process particularly sensitive to the initial parameter settings\cite{jian2010robust}.

To surmount the aforementioned difficulties, this article introduces \SwdFold, which leverages the theory of optimal transport. Optimal transport theory provides a robust mathematical foundation for comparing different data distributions and possesses extensive applicative value across various disciplines, including computer vision, machine learning, and statistical inference\cite{torres2021survey, santambrogio2015optimal}.
It has also been validated as an effective metric and loss function in high-energy physics experiments for measuring the distributional differences between diverse high-dimensional data sets\cite{pan2023event,howard2022learning, fraser2022challenges, cai2020linearized}.
Our \SwdFold facilitates a gradual approximation to the true data distribution without the need for data binning. This method does not rely on any prior physical models during the unfolding process, enhancing its versatility and applicability. On the optimization front, \SwdFold offers a more stable energy landscape for unfolding, which aids in pursuing an optimal solution more effectively. Compared to binary weighted classification reweighting  based algorithms, it is less susceptible to becoming trapped in local optima, thereby ensuring more accurate and robust unfolding results\cite{kolouri2018sliced}.

In exploring the estimation of probability density ratios via \SwdFold, we begin with the fundamental elements of one-dimensional probability measures, transitioning gradually to practical computational domains. The core of this study is the construction of the Cumulative Distribution Function (CDF) on the set of real numbers, $\mathbb{R}$, for a probability measure $\alpha$\cite{deisenroth2020mathematics}. 
The mathematical definition is the integration of the measure $\alpha$ up to a certain point $x \in \mathbb{R}$, expressed as:

\begin{equation}
\forall x \in \mathbb{R}, \quad \mathcal{CDF}_\alpha(x) \stackrel{\text { def. }}{=} \int_{-\infty}^x \mathrm{~d} \alpha .
\end{equation}

For the 1-Wasserstein distance, its computation becomes the focal point of our methodological exposition \cite{kolouri2017optimal}. Against two probability density functions $\alpha, \beta$, its expression is:

\begin{equation}
WD(\alpha, \beta)=\int_{- \infty}^{\infty}\left|\mathcal{CDF}_\alpha(x)-\mathcal{CDF}_\beta(x)\right| \mathrm{~d} x 
\end{equation}

This expression represents the essence of optimal transport, quantifying the work required to transform one probability distribution into another.

For discrete cases, data vectors $A$ and $B$ should be sorted in ascending order. Firstly, we then transfer their weights $w_A$ and $w_B$ into same Range:

\begin{equation}
w'_b = \frac {w_b \sum w_a} {\sum w_b}
\end{equation}

Subsequently, we define the weighted cumulative distribution functions:

\begin{equation}
\mathcal{CDF}_a(x_i)=\sum_{j=1}^i w_a(x_j) , \mathcal{CDF}_b(x_i)=\sum_{j=1}^i w'_b(x_j)
\end{equation}

where $i=1,2, \ldots, n$ and $j=1,2, \ldots. m, n$ and $m$ are the lengths of $A$ and $B$ respectively.

Finally, in the Wasserstein distance computation phase, we calculate the Wasserstein-1 distance as follows:

\begin{equation}
WD\left(A, W_A ; B, W_B\right)=\sum_{i=0}^{n+m-1}\left|\mathcal{C D F}_a\left(x_i\right)-\mathcal{C D F}_b\left(x_i\right)\right| \cdot\left(x_{i+1}-x_i\right)
\end{equation}

where $x \in\left(\mathcal{A} \cup \mathcal{B}\right), x_0 \leq x_1 \leq \ldots \leq x_i \leq \ldots \leq x_{m+n}$.

Thus, we complete the computation of the one-dimensional Wasserstein distance\cite{peyre2019computational_ot}.

\begin{figure}[htbp]
    \centering
    \begin{minipage}{0.48\textwidth}
        \centering
        \includegraphics[width=\linewidth]{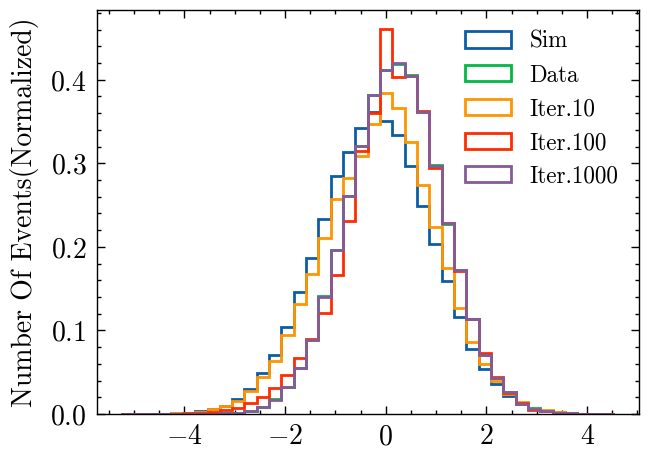}
    \end{minipage}
    \hfill
    \begin{minipage}{0.48\textwidth}
        \centering
        \includegraphics[width=\linewidth]{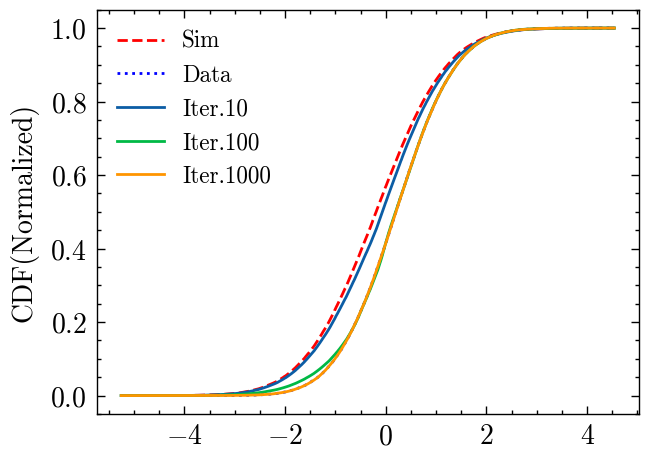}
    \end{minipage}
    \caption{Example for Reweighted Simulation Data Across Iterations using SWDRW.  This figure consists of two plots that illustrate the effectiveness of iterative reweighting of simulated data(Sim) to match actual observed data(Data). The left plot data (Sim) and actual data (Data), alongside reweighted simulated data at various iterations: 10 (Iter.10), 100 (Iter.100), and 1000 (Iter.1000). The right plot is a Cumulative Distribution Function (CDF) over the same value range, comparing the cumulative probability distributions for the same categories of data. These plots are essential for evaluating how effectively the iterative reweighting process adjusts the simulated data to align with the distribution of the actual data, demonstrating increased alignment as the number of iterations grows.}
    \label{fig:SWDRW}
\end{figure}

Next, we extend to the multidimensional case. As Fig.\ref{fig:SWDRW} shown, We adopt a slicing approach, performing slices like radon trasform across different feature quantities in each possible dimension $A_{\theta_1}, A_{\theta_2}, \ldots$ \cite{radon_transfrom, swd_theory}. In practice, we can use huge dimensional like $n=256$ to approximate. Thus We can get

\begin{equation}
SWD\left(A, W_A ; B, W_B\right)= \frac 1 n \sum_{i=1}^n WD(A_{\theta_i}, W_A; B_{\theta_i}, W_B)
\end{equation}

So our optimization objective is to ensure that the estimated probability density ratios are approximately consistent with those derived from optimal transport theory across each possible dimension. This allows us to correctly estimate the probability density ratio in high-dimensional cases. The optimization function can be written as:

\begin{equation}
f_{SWDRW}\left(A, W_A ; B, W_B\right) = {\arg \min}_{W_{ratio}} SWD(A, W_A \cdot W_{ratio}; B, W_B)
\end{equation}

After optimization by gradient descent using Adam optimizer\cite{kingma2014adam}, we can get the probability density ratio between two weighted samples $(A, W_A)$ and $(B, W_B)$ by 

\begin{equation}
f_{SWDRW}\left(A, W_A ; B, W_B\right)(x) = \frac {p_{(B, W_B)}(x)} {p_{(A, W_A)}(x)}
\end{equation}

where $p_{(X, W)}(x)$ means the probability density of x estimated by sample $X$ with weight $W$. Figure \ref{fig:SWDRW} effectively demonstrates this convergence by presenting both a normalized histogram and a cumulative distribution function using two toy normal distribution's samples with different parameters. These visual comparisons showcase how the reweighted simulations iteratively approach the actual data distribution. As indicated in the plots, the successive iterations (10, 100, and 1000) yield increasingly accurate approximations to the true data distribution. This progressive improvement, evident in the narrowing differences between two samples, highlights the efficacy of our optimization method in achieving precise reweighting toward the observed distribution.

To demonstrate the flexibility and efficacy of our method, our study generate toy Monte Carlo events that simulated the studied production and decay processes in phase space. We simulate the particle level and detector level event produced by electron-positron collisions, with a focus on the decay of the $\psi(2S)$ resonance into a $\phi$ meson and a pair of $\pi^+$, $\pi^-$, followed by the further decay of the $\phi$ meson into a pair of $K^+$, $K^-$.  We retained relevant physical information in the phase space Monte Carlo, referred to as synthetic data: one at the particle level, termed "Generation" representing theoretical predictions without detector effects in phase space; and another at the simulation level, termed "Simulation" including simulated detector responses. A straightforward physical model was used to assign weights to the generated data. These weights adjusted the probability distribution of our simulated events to resemble that of another toy Monte Carlo with a physically meaningful probability distribution. Subsequently, we resampled the weighted generated data to produce an unweighted dataset, which we refer to as nature. This nature dataset is markedly different from the synthetic distribution, particularly notable in the probability distribution of some key physical features. Similarly, nature data at the particle level is referred to as “Truth” and at the simulation level as “Data” The unfolding process uses the Simulation, Generation, and Data datasets as inputs, training an algorithm capable of approximating the Truth probability distribution. The objective is to correct experimental effects present in the data, thereby yielding a distribution very close to Truth. Our method allows for the simultaneous unfolding of multiple observables related to the decay process, providing a comprehensive understanding of the underlying physics. Additionally, having weight labels enables us to more accurately compare the performance of various methods and ascertain the accuracy of their estimated distributions.

In the specific iterative unfolding process of \SwdFold, we split into the reweight process(\SwdRw) to learn detector level weights $w_n$ and particle level model learning process to learn particle level weights $v_n$ \cite{andreassen2020omnifold} by

\begin{equation}
f_{Model}(A, 1; A, W_{RW}) = {\arg \max}_g w_{RW} log(g(a)) 
\end{equation}

In the first interation of reweight process, we set particle level weight $\nu_0$ as uniform weights, then we can optimize by the following step:

\begin{equation}
    w_n(x)=\nu_{n-1}(x) f_{\text{SWDRW}}(\text{Sim}, \nu_{n-1}; \text{Data}, 1)(x)
\end{equation}
\begin{equation}
    \nu_{n}(y) = \nu_{n-1}(y)f_{\text{Model}}(\text{Gen}, 1; \text{Gen}, w_n)(y)
\end{equation}

where $(x, y)$ is paired event in synthetic dataset, where x is detector level event, and y is corresponding particle level event. Then after n iterations, we can estimate the unfolded weights as 

\begin{equation}
    p_{(\text{Truth}, 1)}^{(n)}(y) \approx \nu_n(y) p_{(\text{Gen}, 1)}(y)
\end{equation}

We employed a multi-layer fully connected neural network architecture incorporating shortcut connections akin to those found in residual network designs. The internal representation of the network is characterized by 256 -dimensional latent space features, allowing for intricate modeling of the underlying structure of the data. This network processes a diverse set of input features, including kinematic properties and particle identities related to decay events. These features encompass the Cartesian momentum components ( $\left.p_x, p_y, p_z\right)$, total energy (E), magnitude of momentum vector $(p)$, transverse momentum ( $p_T$ ), and angular coordinates ( $\theta$ and $\phi$ ) for $K^+, K^-$, $\pi^+, \pi^-, \phi$, and $\pi^+\pi^-$ pairs. Additionally, the model is sensitive to the involved particle types, namely $\mathrm{K}^+, \mathrm{K}^-, \pi^+$, and $\pi^-$. To further enrich the feature space, the model also incorporates several key invariant mass combinations, particularly the $M_{\phi}$ of the $\phi$ meson and $M_{\pi^+\pi^-}$ for the $\pi^+\pi^-$ pair. Our model was trained using the PyTorch library, and the optimization of network parameters was conducted using the AdamW optimizer, a variant of the Adam optimizer that includes a decoupled weight decay regularization \cite{loshchilov2017decoupled}.

\begin{figure}[H]
    \centering
    \includegraphics[width=0.8\textwidth]{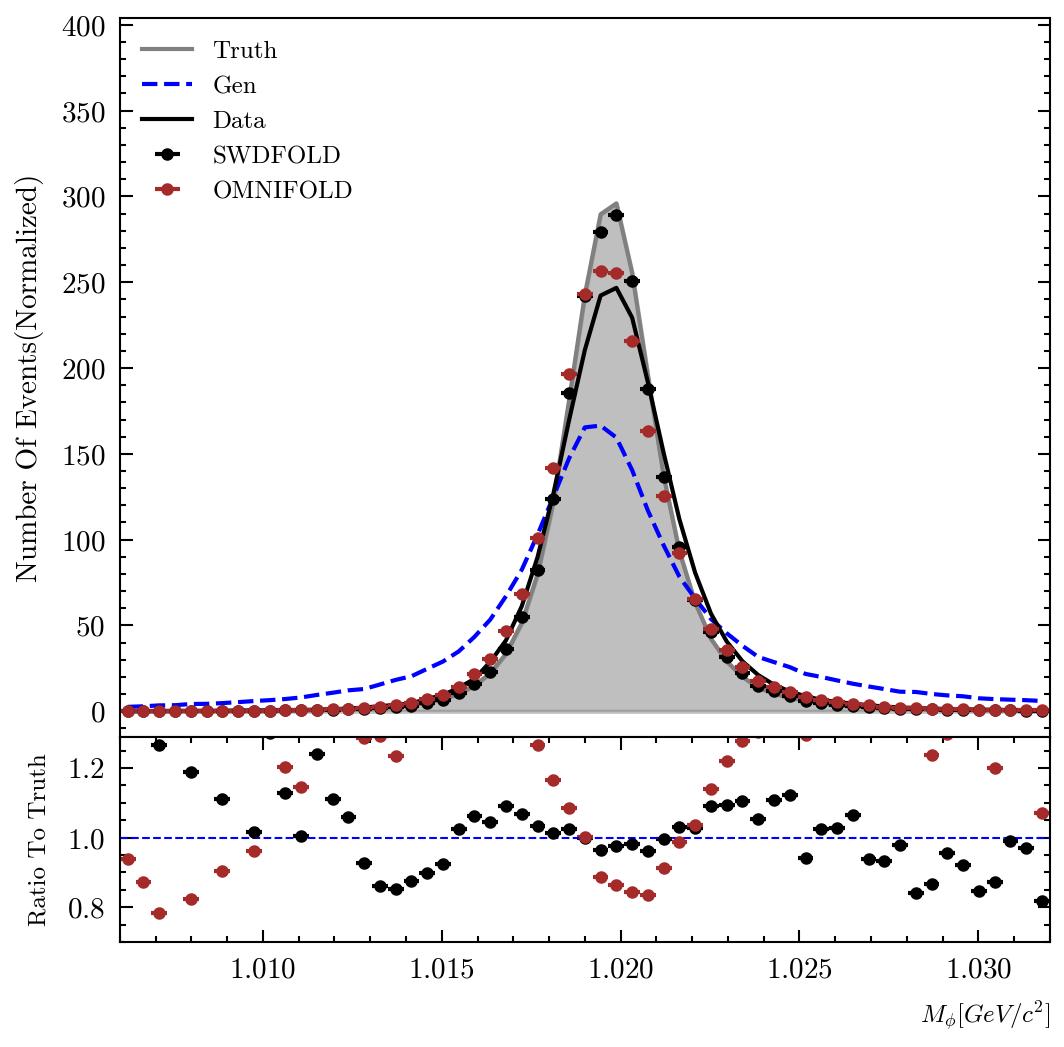}
    \caption{Histogram Analysis of Unfolding Methodologies. The performance of our proposed \SwdFold method is shown with brown dots, compared with the \OmniFold method with black dots. Our \SwdFold method exhibits a high degree of congruence with the truth event(Truth), signifying a successful reweighting of the generation event distribution(Gen) to estimate the truth event distribution(Truth), as particularly evident in the sharp peak around 1.020 GeV/c² in the plot.}
    \label{fig:phi mass}
\end{figure}

Our comprehensive analysis underscores the efficacy of the \SwdFold method, particularly in the context of unfolding and reweighting within our challenging particle physics datasets. The unfolding task in our study was confronted with the intricate difficulty of subtle detector effects, which typically do not manifest in variable distributions when we unfold all the observable in Fig.\ref{fig:all_observables}. This issue was most pronounced when examining the resolution of the $\phi$ mass, where a comparison of detector-level events (Data) and particle-level events (Truth) revealed minimal apparent discrepancies. Nonetheless, as Fig.\ref{fig:phi mass} shown, \SwdFold demonstrated its exceptional capacity to refine the $\phi$ mass distribution, aligning it more closely with the truth distribution and thereby enabling a more precise measurement of the $\phi$ mass width.

\begin{figure}[H]
    \centering
    \begin{subfigure}[b]{0.9\textwidth}
        \includegraphics[width=\textwidth]{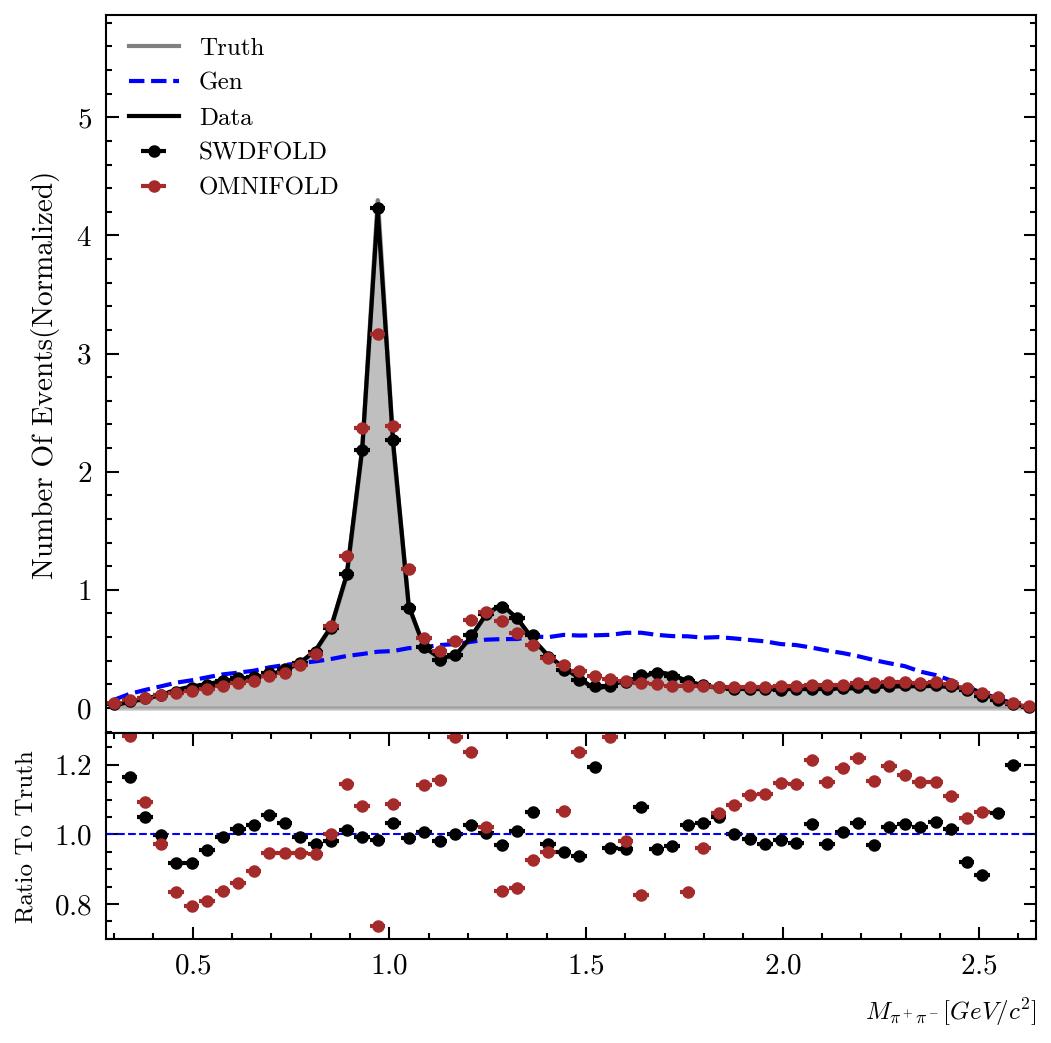}
    \end{subfigure}
    \hfill
    \begin{subfigure}[b]{0.3\textwidth}
        \includegraphics[width=\textwidth]{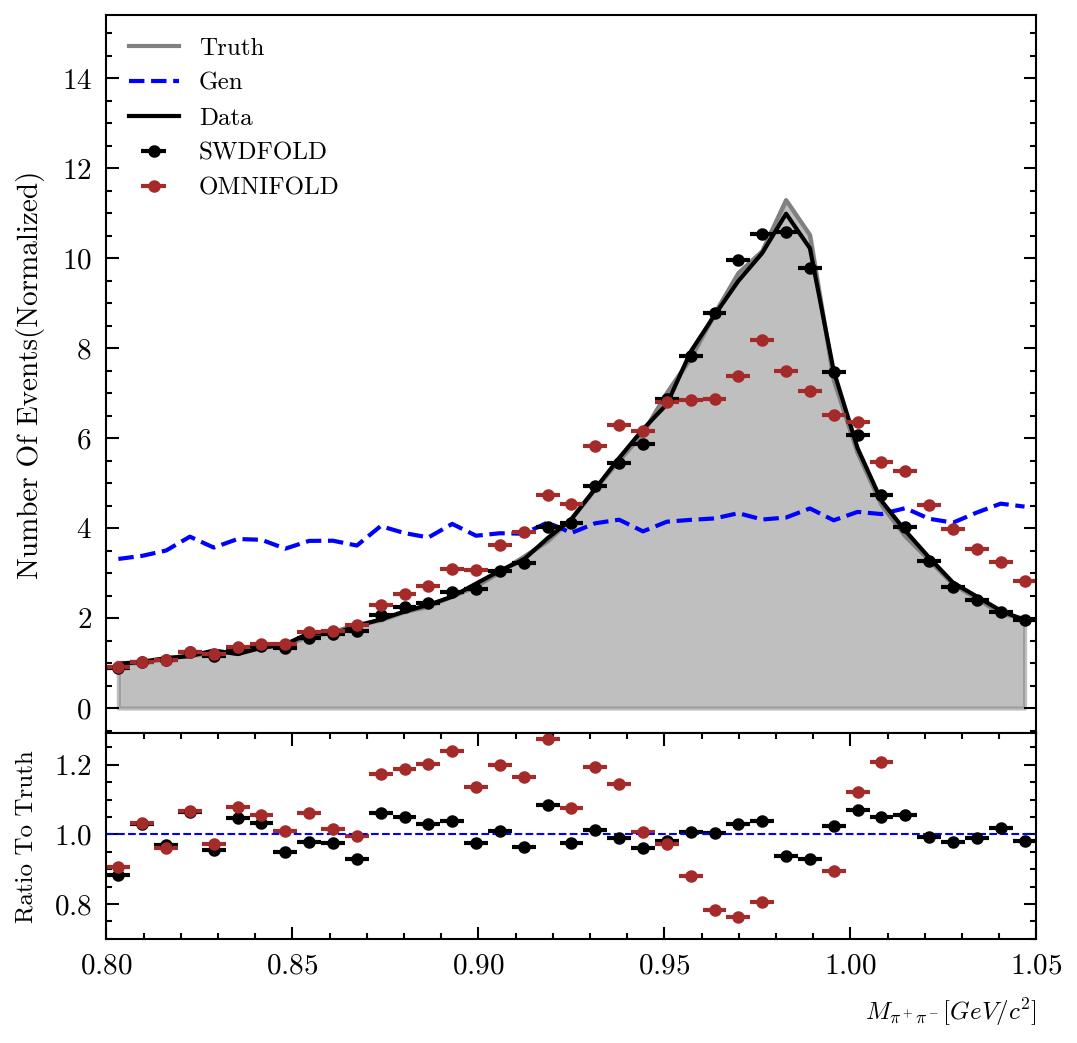}
    \end{subfigure}
    \begin{subfigure}[b]{0.3\textwidth}
        \includegraphics[width=\textwidth]{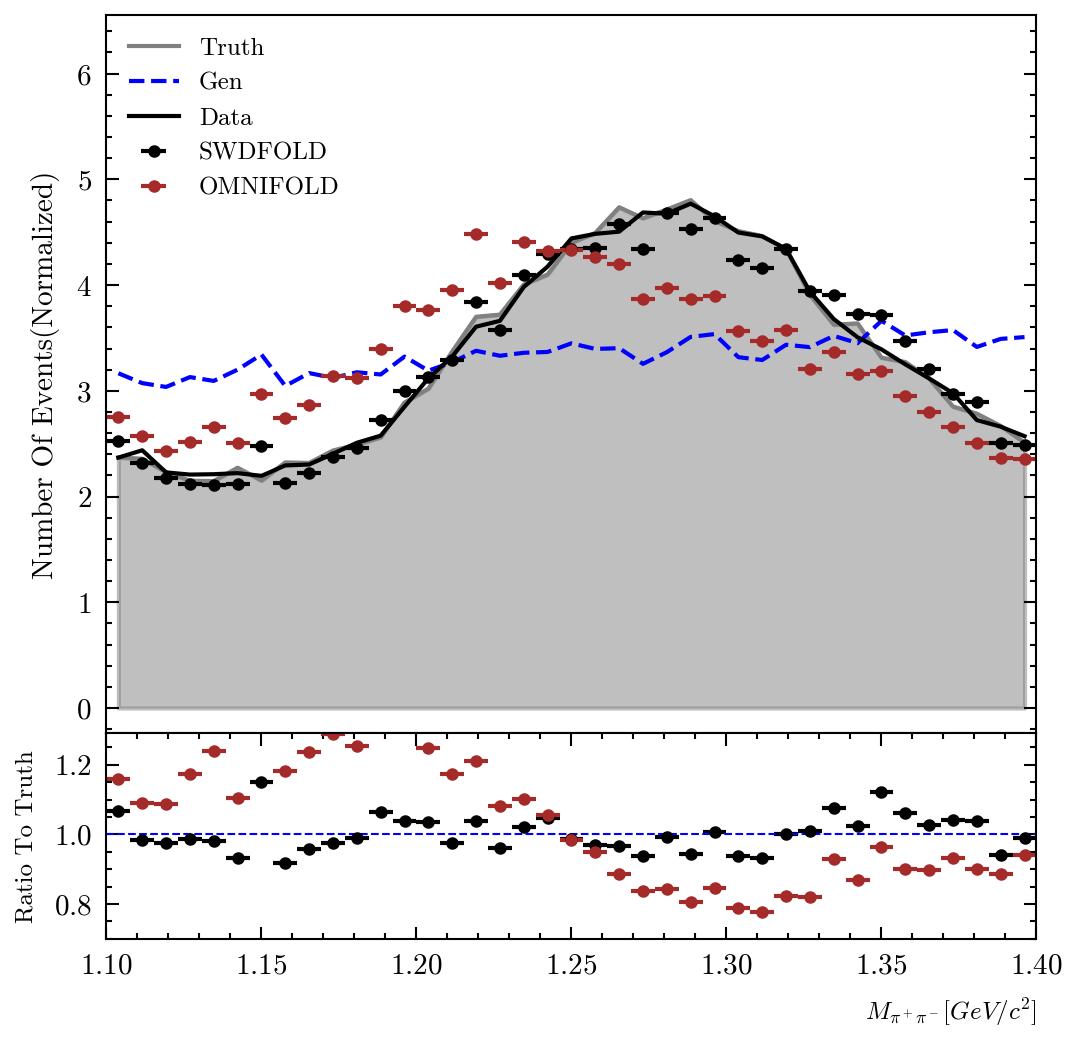}
    \end{subfigure}
     \begin{subfigure}[b]{0.3\textwidth}
        \includegraphics[width=\textwidth]{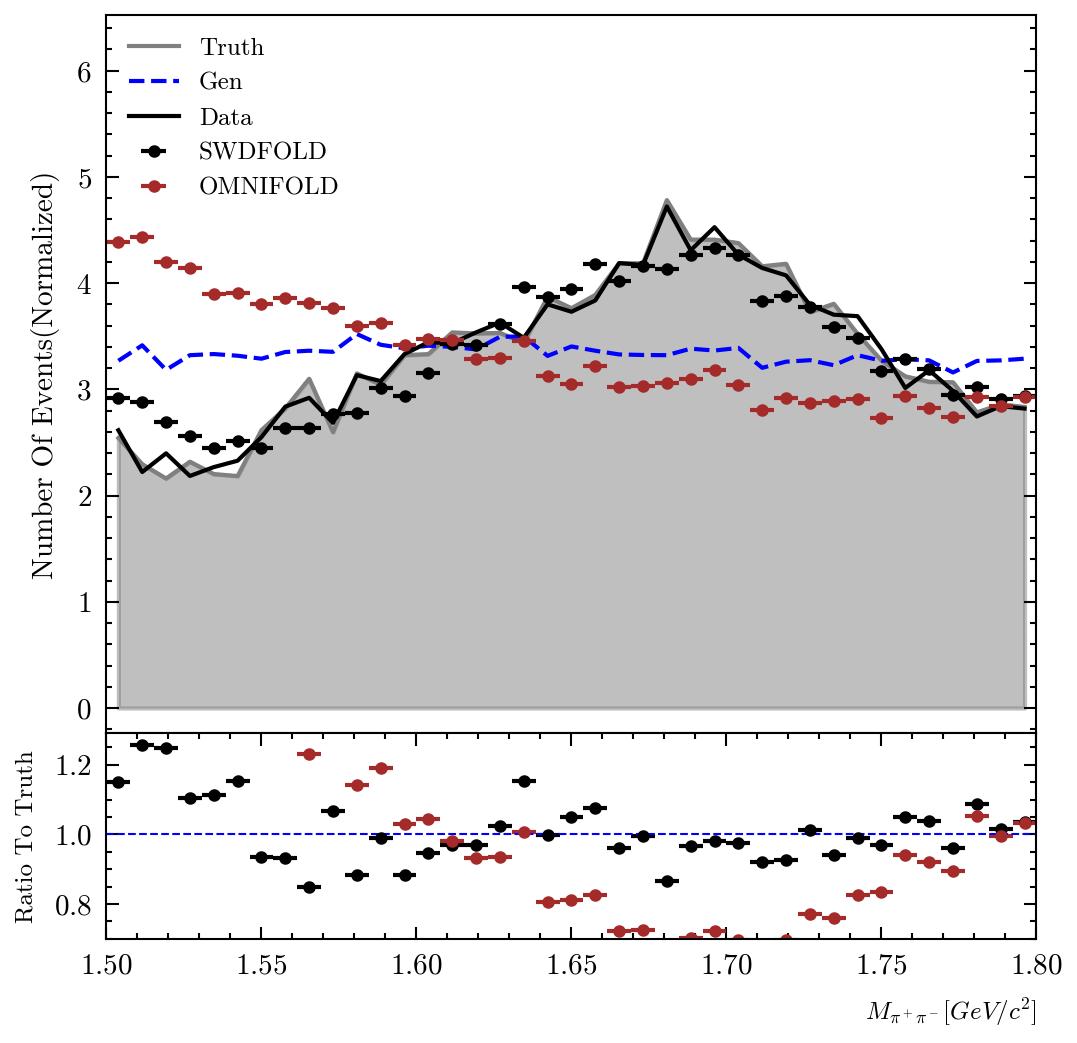}
    \end{subfigure}

    \caption{Histogram Analysis of Reweight Methodologies. We presents a detailed examination of the $\pi^{+} \pi^{-}$mass distribution across three distinct mass ranges: $(0.8,1.05),(1.1,1.4)$, and $(1.5,1.8) \mathrm{GeV} / \mathrm{c}^2$. The performance of the \SwdFold method, illustrated by the black line, showcases its ability to reveal the resonance peaks by reweighting the Generation Event(Gen), aligning closely with the Truth Event(Truth) in both peak value and width.}
    \label{fig:pipi mass}
\end{figure}

The reweighting process faced its own set of challenges, especially when comparing synthetic datasets with natural datasets. In the natural dataset, the $\pi^{+} \pi^{-}$mass distribution displayed pronounced peaks, signaling resonance in decay-peaks that were missing from the synthetic dataset. \SwdFold's robust algorithmic framework not only succeeded in accurately representing these resonant peaks but also replicated the corresponding widths observed in nature in Fig.\ref{fig:pipi mass}. These advancements in reweighting signify \SwdFold's nuanced capability to accurately reflect complex physical phenomena, thereby enhancing data analysis and parameter estimation within the domain of particle physics research.

\begin{figure}[H]
  \centering
  
  \begin{subfigure}{.32\textwidth}
    \centering
    \includegraphics[width=.9\linewidth]{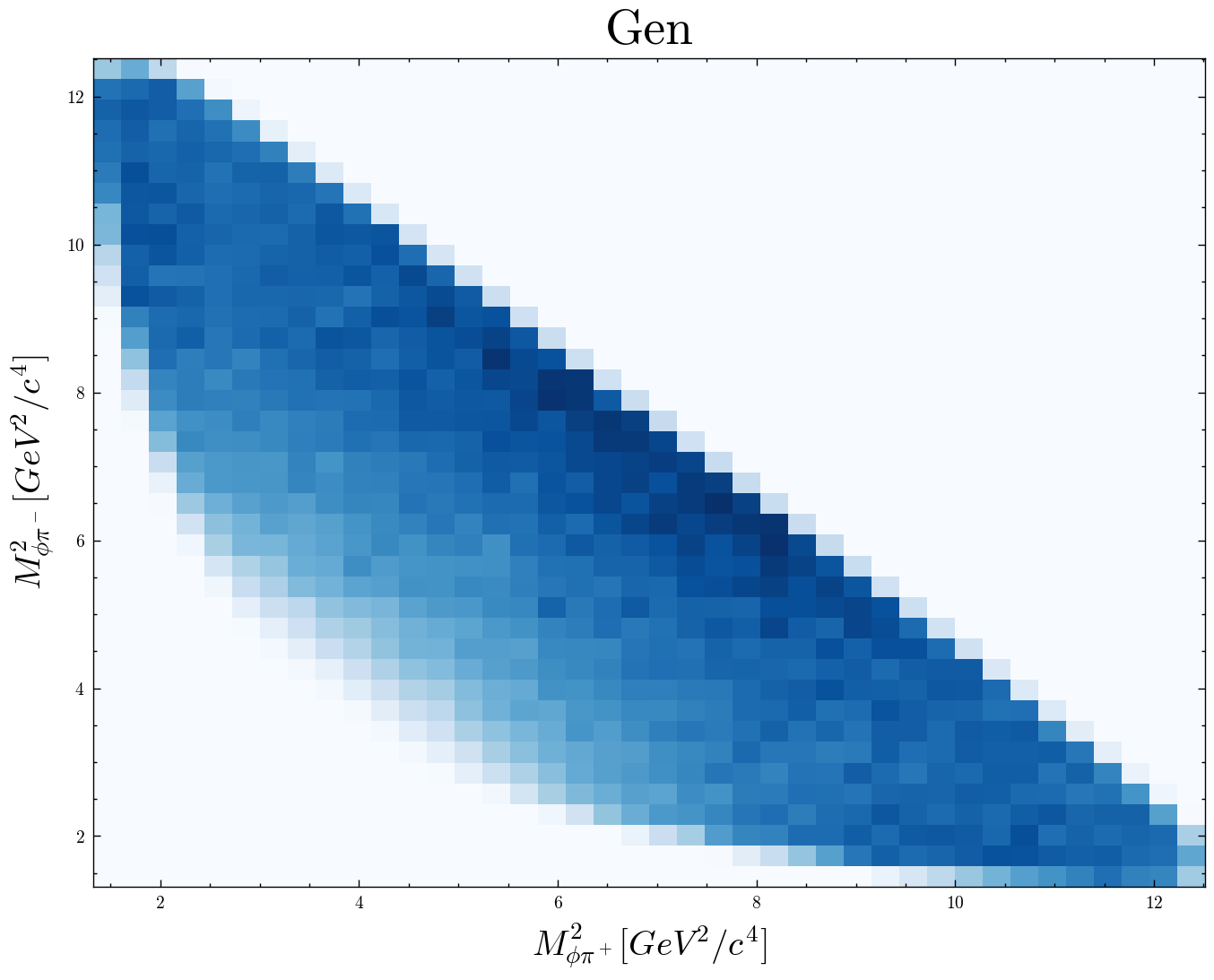}
  \end{subfigure}
  \hfill
  \begin{subfigure}{.32\textwidth}
    \centering
    \includegraphics[width=.9\linewidth]{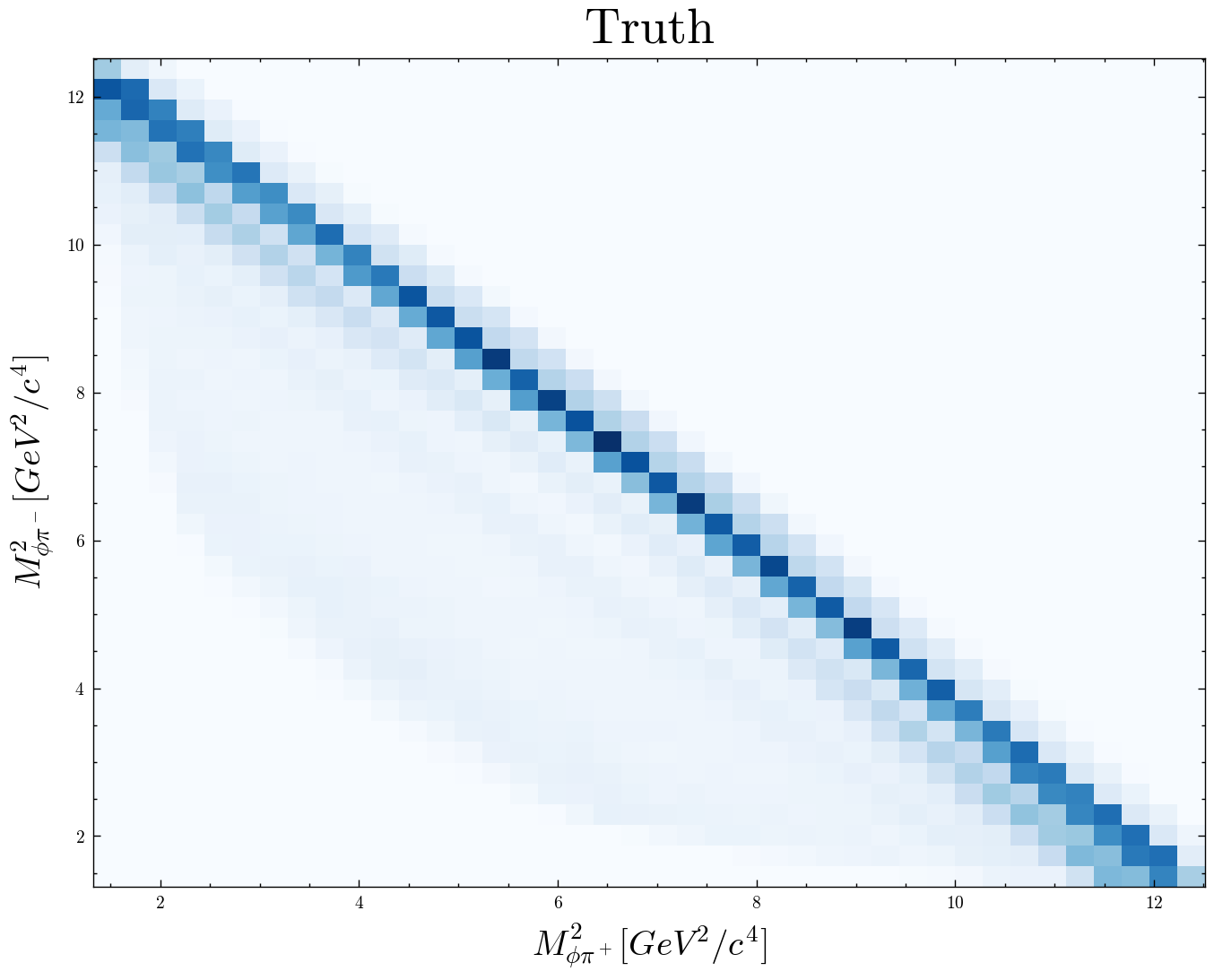}
  \end{subfigure}
  \hfill
  \begin{subfigure}{.32\textwidth}
    \centering
    \includegraphics[width=.9\linewidth]{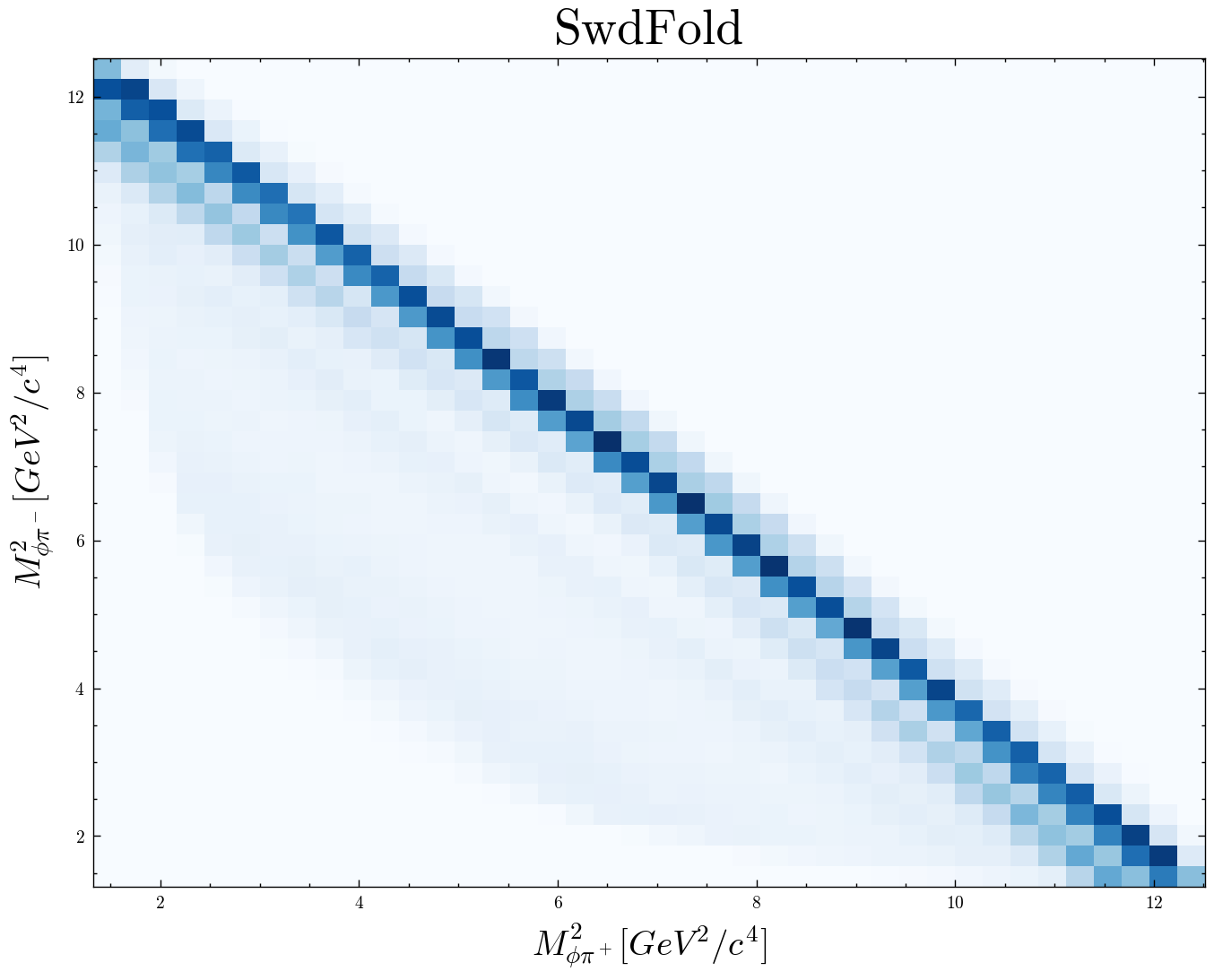}
  \end{subfigure}
  
  \caption{Visualization of the Dalitz plots for the $\phi \pi$ system. The left panel depicts the distribution from the generated dataset (Gen), which serves as the simulated data before any reweighting. The center panel represents the Truth data, reflecting the actual particle interactions without the influence of detector effects or experimental noise. The right panel showcases the distribution post the application of the SwdFold method. The SwdFold plot is close to the Truth distribution, demonstrating its successful reweighting.}
  \label{fig:dalitz plot}
  \end{figure}

Our \SwdFold method not only achieves high agreement with the one-dimensional distributions, has demonstrated remarkable success in capturing and preserving the intricate correlations present within the multi-dimensional phase space of particle interactions. As evidenced by the Dalitz plots presented in Fig.\ref{fig:dalitz plot}, the close alignment of the \SwdFold plot with the Truth data attests to the method's precision in reweighting and its adeptness in handling the complex correlations characteristic of particle physics data. This proficiency extends beyond mere visual congruence; it indicates a deeper, quantitative concordance across the full kinematic domain, showcasing the potential of our method for high-precision reweighting and parameter estimation. The ability of \SwdFold to deliver physically meaningful results, positions it as a powerful tool for enhancing the accuracy and reliability of analyses in experimental particle physics.

In conclusion, we presents \SwdFold, a pioneering reweighting and unfolding method grounded in the principles of optimal transport, which addresses the longstanding challenges faced by traditional unfolding techniques in high-energy physics. Our method significantly mitigates issues related to model dependency, binning artifacts, and sensitivity to statistical fluctuations, which are prevalent in conventional methods. By deploying sliced wasserstein distance theory to approximate the probability density ratio of simulation dataset and experimental dataset, \SwdFold can reweight the particle level simulation event to estimate the true physical result without detector effect. The empirical validations using high-energy physics datasets, specifically the simulated events of electron-positron collisions and the decay processes $\psi(2S) \rightarrow \phi K^+ K^-, \phi \rightarrow K^+ K^-$, demonstrate that \SwdFold can accurately recover the true distribution of physical events with remarkable precision. Notably, the method excels in maintaining the integrity of multidimensional correlations and nuanced physical features, which are often obscured by detector effects and other experimental noise. \SwdFold's innovative approach to data unfolding and its ability to handle high-dimensional data without binning represent substantial advancements in the field. These enhancements enable physicists to derive more accurate insights from experimental data, thereby facilitating a deeper understanding of underlying physical phenomena and contributing to the ongoing development of theoretical models. Future work will focus on refining the algorithm's efficiency and exploring its application in other complex systems where data integrity is critical. The adaptability and performance of \SwdFold affirm its potential as a valuable tool for a wide range of applications in experimental physics and beyond, promising to significantly impact how data-driven discoveries are made in the sciences.

Looking ahead, \SwdFold paves new possibilities for directly training deep learning models in an unsupervised manner for tasks in unfolding and reweighting. This approach holds promising prospects not only for future research in particle physics. To foster continued research and development in the field, we plan to make our \SwdFold code and related datasets publicly available. This initiative will enable other researchers to apply our method in various settings, further validating and enhancing its applicability.

\bibliographystyle{unsrtnat}
\bibliography{references}

\appendix
\section{Additional Figure}

\begin{figure}[!ht]
    \centering
    \foreach \n in {0,..., 19} {%
      \begin{subfigure}{0.22\textwidth}
        \includegraphics[width=\linewidth]{fig/unfold_compare_fig/\n.jpg}
      \end{subfigure}%
      \hfill 
    }
    \end{figure}
    
    \begin{figure}[!ht]
    \centering
    \foreach \n in {20,...,39} {%
      \begin{subfigure}{0.22\textwidth}
        \includegraphics[width=\linewidth]{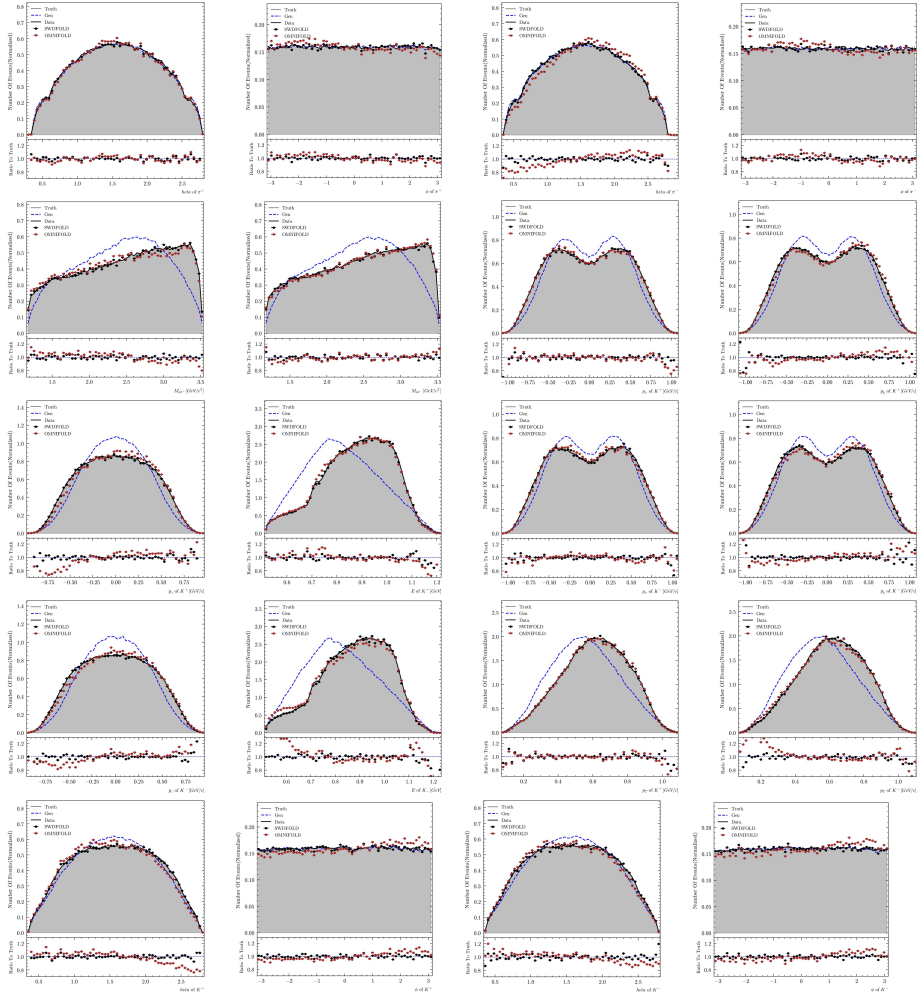}
      \end{subfigure}%
      \hfill
    }
    \caption{The unfolding result for all observables}
    \label{fig:all_observables}
    \end{figure}

\end{document}